 \documentstyle[preprint,tighten,eqsecnum,prc,aps]{revtex}
\begin{document}
\hyphenation{Nijmegen}
\hyphenation{Rijken}
\draft
\preprint{\vbox{. \hfill ADP-99-3-T348}}

\title{Soft-core baryon-baryon potentials for the complete baryon octet}
\author{V.G.J.\ Stoks}
\address{Physics Division, Argonne National Laboratory, Argonne,
         Illinois 60439 \protect\\ and
         Centre for the Subatomic Structure of Matter,
         University of Adelaide, Adelaide, SA 5005, Australia}
\author{Th.A.\ Rijken}
\address{Institute for Theoretical Physics, University of Nijmegen,
         Nijmegen, The Netherlands}
\date{}
\maketitle

\begin{abstract}
SU(3) symmetry relations on the recently constructed hyperon-nucleon
potentials are used to develop potential models for all possible
baryon-baryon interaction channels. The main focus is on the interaction
channels with total strangeness $S=-2$, $-3$, and $-4$, for which no
experimental data exist yet. The potential models for these channels are
based on SU(3) extensions of potential models for the $S=0$ and $S=-1$
sectors, which {\it are\/} fitted to experimental data. Although the
SU(3) symmetry is not taken to be exact, the $S=0$ and $S=-1$ sectors
still provide the necessary constraints to fix all free parameters.
The potentials for the $S=-2$, $-3$, and $-4$ sectors, therefore,
do not contain any additional free parameters, which makes them
the first models of this kind.
Various properties of the potentials are illustrated by giving results
for scattering lengths, bound states, and total cross sections.
\end{abstract}

\narrowtext

\section{Introduction}
The study of strangeness-rich systems is of fundamental importance
in understanding relativistic heavy-ion collisions~\cite{Shi95},
some astrophysical problems~\cite{Bay95,Pra98}, and the existence
(or nonexistence) of certain hypernuclei.
Strangeness-rich systems can be exotic multiquark systems consisting
of up ($u$), down ($d$), and strange ($s$) quarks; like the elusive
$H$ dibaryon, a 6-quark $uuddss$ system predicted by Jaffe~\cite{Jaf77}.
But they can also simply be bound states of nucleons ($N$), hyperons
($Y=\Lambda,\Sigma$), and cascades ($\Xi$). In order to get a better
handle on the latter possibility, we are in need of potential models
which describe all possible interactions between nucleons, hyperons,
and cascades.

Although there is a wealth of accurate $N\!N$ scattering data, which
allows us to construct accurate $N\!N$ potential models, there are only
a few $Y\!N$ scattering data, and there are no scattering data at all
for the multi-strange systems $YY$, $\Xi N$, $\Xi Y$, and $\Xi\Xi$.
This means that the potential models for these multi-strange
interactions require input from elsewhere to define them.
One possibility is to use experimental information on bound states
of hypernuclei. Double-$\Lambda$ hypernuclei, for example, provide
information on the $\Lambda\Lambda$ and $\Xi N$ interactions. 
However, here one has to be careful since the extracted information
is, in a sense, ``contaminated'' by many-body effects. Furthermore,
there are no hypernuclear experimental data yet which could provide
information on the $\Xi Y$ or $\Xi\Xi$ interactions.

In this paper, we therefore consider a second possibility, which is
to assume that the potentials obey a (slightly broken) SU(3) symmetry.
The potentials are parametrized in terms of one-boson exchanges,
which we believe to be a very good and, certainly, effective first
approximation in modeling the complete interaction. Extensions beyond
the one-meson-exchange mechanism, like the inclusion of two-meson
exchanges and $\Delta$ and $Y^*$ isobars in intermediate states,
are expected to be of lesser importance. The assumption of SU(3)
symmetry allows us to determine all coupling constants in a fit to
the $N\!N$ and $Y\!N$ scattering data, which also defines all the
coupling constants needed to describe the multi-strange interactions.
However, the fit to the $N\!N$ and $Y\!N$ data still allows for some
freedom in the parameters, and so in Ref.~\cite{Rij99} we have
constructed six different $Y\!N$ models. The different models are
characterized by different choices for the magnetic vector $F/(F+D)$
ratio, $\alpha_V^m$, which serves to produce different scattering
lengths in the $\Lambda N$ and $\Sigma N$ channels, but at the same
time allows all models to describe the available $Y\!N$ (and $N\!N$)
scattering data equally well. The values chosen for $\alpha_V^m$
range from 0.4447 (model NSC97a) to 0.3647 (model NSC97f).
Within each model, there are now no free parameters left, and so
each parameter set defines a baryon-baryon potential which models
all possible two-baryon interactions.

Although most of the details on the $N\!N$ and $Y\!N$ interactions are
well-known and can be found elsewhere, we have here decided to include
them in order to present a complete picture of how our baryon-baryon
potentials are defined. Therefore, in Sec.~\ref{sec:channels} we
first present the SU(3)-symmetric interaction Lagrangian describing
the interaction vertices between mesons and members of the
$J^P={\textstyle\frac{1}{2}}^+$ baryon octet, and define their coupling
constants. (As stated above, states involving the members of the
$J^P={\textstyle\frac{3}{2}}^+$ baryon decuplet are expected to be
of lesser importance, and their inclusion is left for a future
investigation.) We then identify the various channels which are
possible for the baryon-baryon interaction. In most cases, the
interaction is a multichannel interaction, characterized by transition
potentials and thresholds. Details are given in Sec.~\ref{sec:trans}.
Together with the appendix of our previous publication~\cite{Rij99},
it is now straightforward to construct the potentials for all the
possible baryon-baryon interaction channels.
In Sec.~\ref{sec:results} we present the general features of the
potentials for all the sectors with total strangeness $S=0,\ldots,-4$.
We give the $S$-wave scattering lengths, discuss the possibility of
bound states in these partial waves, and give results for the total
cross sections for all leading channels.
We conclude with Sec.~\ref{sec:conc}.

\section{Baryon-baryon channels}
\label{sec:channels}
We consider all possible baryon-baryon interaction channels, where
the baryons are the members of the $J^P={\textstyle\frac{1}{2}}^+$
baryon octet
\begin{equation}
  B = \left( \begin{array}{ccc}
      {\displaystyle\frac{\Sigma^{0}}{\sqrt{2}}+\frac{\Lambda}{\sqrt{6}}}
               &  \Sigma^{+}  &  p  \\[2mm]
      \Sigma^{-} & {\displaystyle-\frac{\Sigma^{0}}{\sqrt{2}}
                   +\frac{\Lambda}{\sqrt{6}}}  &  n \\[2mm]
      -\Xi^{-} & \Xi^{0} &  {\displaystyle-\frac{2\Lambda}{\sqrt{6}}}
             \end{array} \right).
\end{equation}
The empirical baryon masses, as quoted by the Particle Data
Group~\cite{PDG96}, are given in Table~\ref{tabmass}.
The meson nonets can be written as
\begin{equation}
     P=P_{\rm sin}+P_{\rm oct},
\end{equation}
where the singlet matrix $P_{\rm sin}$ has elements $\eta_0/\sqrt{3}$
on the diagonal, and the octet matrix $P_{\rm oct}$ is given by
\begin{equation}
   P_{\rm oct} = \left( \begin{array}{ccc}
      {\displaystyle\frac{\pi^{0}}{\sqrt{2}}+\frac{\eta_{8}}{\sqrt{6}}}
             & \pi^{+}  &  K^{+}  \\[2mm]
      \pi^{-} & {\displaystyle-\frac{\pi^{0}}{\sqrt{2}}
         +\frac{\eta_{8}}{\sqrt{6}}}  &   K^{0} \\[2mm]
      K^{-}  &  \overline{{K}^{0}}
             &  {\displaystyle-\frac{2\eta_{8}}{\sqrt{6}}}
             \end{array} \right),
\end{equation}
and where we took the pseudoscalar mesons with $J^P=0^+$ as a specific
example. Introducing the following notation for the isodoublets,
\begin{equation}
  N=\left(\begin{array}{c} p \\ n \end{array} \right), \ \ \
  \Xi=\left(\begin{array}{c} \Xi^{0} \\ \Xi^{-} \end{array} \right), \ \ \
  K=\left(\begin{array}{c} K^{+} \\ K^{0} \end{array} \right),
  \ \ \   K_{c}=\left(\begin{array}{c} \overline{K^{0}} \\
               -K^{-} \end{array} \right),        \label{doublets}
\end{equation}
the most general, SU(3) invariant, interaction Lagrangian is then
given by~\cite{Swa63}
\begin{eqnarray}
   m_{\pi}{\cal L}_{\rm pv}^{\rm oct} &=&
  -f_{N\!N\pi}(\overline{N}\bbox{\tau}N)\!\cdot\!\bbox{\pi}
  +if_{\Sigma\Sigma\pi}(\overline{\bbox{\Sigma}}\!\times\!\bbox{\Sigma})
      \!\cdot\!\bbox{\pi}
  -f_{\Lambda\Sigma\pi}(\overline{\Lambda}\bbox{\Sigma}+
      \overline{\bbox{\Sigma}}\Lambda)\!\cdot\!\bbox{\pi}
  -f_{\Xi\Xi\pi}(\overline{\Xi}\bbox{\tau}\Xi)\!\cdot\!\bbox{\pi}
            \nonumber\\
 &&-f_{\Lambda N\!K}\left[(\overline{N}K)\Lambda
         +\overline{\Lambda}(\overline{K}N)\right]
   -f_{\Xi\Lambda K}\left[(\overline{\Xi}K_{c})\Lambda
         +\overline{\Lambda}(\overline{K_{c}}\Xi)\right] \nonumber\\
 &&-f_{\Sigma N\!K}\left[\overline{\bbox{\Sigma}}\!\cdot\!
         (\overline{K}\bbox{\tau}N)+(\overline{N}\bbox{\tau}K)
         \!\cdot\!\bbox{\Sigma}\right]
   -f_{\Xi\Sigma K}\left[\overline{\bbox{\Sigma}}\!\cdot\!
       (\overline{K_{c}}\bbox{\tau}\Xi)
     +(\overline{\Xi}\bbox{\tau}K_{c})\!\cdot\!\bbox{\Sigma}\right]
                                         \nonumber\\
 &&-f_{N\!N\eta_{8}}(\overline{N}N)\eta_{8}
   -f_{\Lambda\Lambda\eta_{8}}(\overline{\Lambda}\Lambda)\eta_{8}
   -f_{\Sigma\Sigma\eta_{8}}(\overline{\bbox{\Sigma}}\!\cdot\!
       \bbox{\Sigma})\eta_{8}
   -f_{\Xi\Xi\eta_{8}}(\overline{\Xi}\Xi)\eta_{8}     \nonumber\\
 &&-f_{N\!N\eta_{0}}(\overline{N}N)\eta_{0}
   -f_{\Lambda\Lambda\eta_{0}}(\overline{\Lambda}\Lambda)\eta_{0}
   -f_{\Sigma\Sigma\eta_{0}}(\overline{\bbox{\Sigma}}\!\cdot\!
       \bbox{\Sigma})\eta_{0}
   -f_{\Xi\Xi\eta_{0}}(\overline{\Xi}\Xi)\eta_{0},  \label{Lint}
\end{eqnarray}
where we again took the pseudoscalar mesons as an example, dropped
the Lorentz character of the interaction vertices (which is
$\gamma_{5}\gamma_{\mu}\partial^{\mu}$ for pseudoscalar mesons),
and introduced the charged-pion mass to make the pseudovector coupling
constant $f$ dimensionless.
All coupling constants can be expressed in terms of only four parameters.
The explicit expressions can be found in Ref.~\cite{Rij99}.
The $\Sigma$-hyperon is an isovector with phase chosen such~\cite{Swa63}
that
\begin{equation}
  \bbox{\Sigma}\!\cdot\!\bbox{\pi} = \Sigma^{+}\pi^{-}
       +\Sigma^{0}\pi^{0}+\Sigma^{-}\pi^{+}.
\end{equation}
This definition for $\Sigma^+$ differs from the standard Condon and
Shortley phase convention~\cite{Con35} by a minus sign. This means that,
in working out the isospin multiplet for each coupling constant in
Eq.~(\ref{Lint}), each $\Sigma^+$ entering or leaving an interaction
vertex has to be assigned an extra minus sign. However, if the potential
is first evaluated on the isospin basis and then, via an isospin
rotation, transformed to the potential on the physical particle basis
(see below), this extra minus sign will be automatically accounted for.

Given the interaction Lagrangian (\ref{Lint}) and a theoretical
scheme for deriving the potential representing a particular Feynman
diagram, it is now straightforward to derive the one-meson-exchange
baryon-baryon potentials.
We follow the Thompson approach~\cite{Tho70,Rij91,Rij92,Rij96} and
expressions for the potential in momentum space can be found in
Ref.~\cite{Mae89}. Unfortunately, the expressions for the potential in
configuration space in this reference contain a number of typographical
errors; the corrected expressions are given in the appendix of our
previous publication on the $Y\!N$ potentials~\cite{Rij99}.

Since the nucleons have strangeness $S=0$, the hyperons $S=-1$,
and the cascades $S=-2$, the possible baryon-baryon interaction
channels can be classified according to their total strangeness,
ranging from $S=0$ for $N\!N$ to $S=-4$ for $\Xi\Xi$.
Apart from the wealth of accurate $N\!N$ scattering data for the total
strangeness $S=0$ sector, there are only a few $Y\!N$ scattering data
for the $S=-1$ sector, while there are no data at all for the $S<-1$
sectors. We therefore believe that at this stage it is not yet
worthwhile to explicitly account for the small mass differences between
the specific charge states of the baryons and mesons; i.e., we use
average masses, isospin is a good quantum number, and the potentials
are calculated on the isospin basis. The possible channels on the
isospin basis are given in Table~\ref{tabisobasis}.

However, the Lippmann-Schwinger or Schr\"odinger equation is solved for
the physical particle channels, and so scattering observables are
calculated using the proper physical baryon masses.
The possible channels on the physical particle basis can be classified
according to the total charge $Q$; these are given in
Table~\ref{tabpartbasis}. The corresponding potentials are obtained
from the potential on the isospin basis by making the appropriate
isospin rotation. The matrix elements of the isospin rotation matrices
are nothing else but the Clebsch-Gordan coefficients for the two baryon
isospins making up the total isospin. (Note that this is the reason why
the potential on the particle basis, obtained from applying an isospin
rotation to the potential on the isospin basis, will have the correct
sign for any coupling constant on a vertex, including the ones
involving a $\Sigma^+$.)

We should point out that this approach does not result in a
high-precision potential for $N\!N$ scattering in the $S=0$ sector.
For one thing, any $N\!N$ potential which is claimed to be of high
precision should at least explicitly account for the fact that the
neutral pion and charged pion have different masses, because one-pion
exchange is the longest-range and most important part of the interaction.
Furthermore, it is well-known that the $^1S_0$ $pp$ and $np$ partial
waves show a sizable breaking of charge independence, which cannot
be explained within a simple one-boson-exchange model like the one
presented here, but requires the inclusion of meson mixing, two-meson
exchange, and/or extra phenomenological parameters. 
The $N\!N$ potential presented here is only included for reasons of
completeness; although it certainly describes the qualitative features
of the baryon-baryon potential for the $J^P={\textstyle\frac{1}{2}}^+$
baryon octet in the $S=0$ sector very well, for high-accuracy
quantitative calculations we recommend that one uses one of the recently
constructed high-precision $N\!N$ models~\cite{Sto94,Wir95,Mac96}.

In order to construct the potentials on the isospin basis, we need
the matrix elements of the various meson exchanges between particular
isospin states. The way to calculate these matrix elements is outlined,
e.g., in Ref.~\cite{Bri68}. The results are given in Table~\ref{tabiso},
where we use the pseudoscalar mesons as a specific example.
The entries also include factors $1/\sqrt{2}$ whenever the initial or
final state consists of two identical particles, and the exchange
operator $P$ for the contributions where the final-state baryons
have been interchanged. The exchange operator $P$ has the value
$P=+1$ for even-$L$ singlet and odd-$L$ triplet partial waves, and
$P=-1$ for odd-$L$ singlet and even-$L$ triplet partial waves.
For total strangeness $S=-1$ and $S=-3$, the final-state interchanged
diagram only occurs when the exchanged meson carries strangeness
($K$, $K^*$, $\kappa$, $K^{**}$). An interesting subtlety is that in
the entry for $(\Xi N|K|\Sigma\Lambda)$ the direct and exchange
contributions carry {\it different\/} coupling constants, which is
the reason why they are not added together, but are given separately.

Finally, in constructing the potentials on the particle basis by
applying the appropriate isospin rotation to the potential on the
isospin basis, care must be taken that in a number of cases the two
initial-state and/or final-state baryons belong to different charge
states within the same isospin multiplet. In those cases, the
multiplication with the identical-particle symmetry factor $1/\sqrt{2}$
has to be undone. In practice, this means that for each initial state
or final state consisting of $|np)$, $|\Sigma^0\Sigma^+)$,
$|\Sigma^-\Sigma^0)$, $|\Sigma^-\Sigma^+)$, or $|\Xi^-\Xi^0)$, the
potential has to be multiplied by $\sqrt{2}$.

We conclude by mentioning that there is nothing that prohibits us
from constructing the potentials directly on the particle basis,
explicitly accounting for all the different charged states of the
baryons and mesons. Although this {\it is\/} the only proper way if
one is to construct a high-precision $N\!N$ potential for the $S=0$
sector, we again note that the scattering data to define the $S=-1$
sector are very scarce, and that there are no scattering data at
all to define the $S<-1$ sectors. Hence, we argue that at this stage
such a refinement does not yet seem to be worth the effort.

\section{Transition potentials}
\label{sec:trans}
\subsection{Thresholds}
The fact that the initial-state and final-state baryons in the
$S=-1$, $-2$, and $-3$ sectors can consist of different baryons
leads to so-called transition potentials. Their presence turns the
Lippmann-Schwinger or Schr\"odinger equation into a coupled-channel
matrix equation where the different channels open up depending on
whether the on-shell energy exceeds a certain threshold.
Let us consider the case where particle 1 with laboratory momentum
$p_{\rm lab}$ scatters off particle 2, which is at rest in the
laboratory frame, and that they represent the lowest-mass two-baryon
state for a specific $(S,Q)$ interaction channel. The total energy
squared is then given by
\begin{equation}
    s=M_1^2(1)+M_2^2(1)+2M_2(1)\sqrt{p_{\rm lab}^2(1)+M_1^2(1)},
\end{equation}
where the 1 in parentheses refers to the fact that we are considering
the lowest-mass (i.e., first channel) two-baryon state.
The center-off-mass (cm) momentum squared in each channel $i$ within
this $(S,Q)$ coupled-channel system is then given by
\begin{equation}
   p^2_{\rm cm}(i)=\frac{1}{4s}\left[s-\{M_1(i)+M_2(i)\}^2\right]
                               \left[s-\{M_1(i)-M_2(i)\}^2\right].
\end{equation}
Clearly, for small values of $p_{\rm lab}(1)$ we find that
$p^2_{\rm cm}(i)$, $(i>1)$, is negative, which means that these
channels are closed. A discussion of how to handle the presence of
closed channels is given, for example, in Ref.~\cite{Nag73}.
The thresholds where the higher-mass channels open up are as follows.
For $(\Lambda p,\Sigma^+n,\Sigma^0p)$:
\begin{eqnarray}
   && p_{\rm lab}^{\rm th}(\Lambda p\rightarrow\Sigma^+n)
                  = 633.4 \mbox{ MeV}/c, \nonumber\\
   && p_{\rm lab}^{\rm th}(\Lambda p\rightarrow\Sigma^0p)
                  = 642.0 \mbox{ MeV}/c.
\end{eqnarray}
For $(\Lambda n,\Sigma^0n,\Sigma^-p)$:
\begin{eqnarray}
   && p_{\rm lab}^{\rm th}(\Lambda n\rightarrow\Sigma^0n)
                  = 641.7 \mbox{ MeV}/c, \nonumber\\
   && p_{\rm lab}^{\rm th}(\Lambda n\rightarrow\Sigma^-p)
                  = 657.9 \mbox{ MeV}/c.
\end{eqnarray}
For $(\Xi^0p,\Sigma^+\Lambda,\Sigma^0\Sigma^+)$:
\begin{eqnarray}
   && p_{\rm lab}^{\rm th}(\Xi^0p\rightarrow\Sigma^+\Lambda)
                  = 589.0 \mbox{ MeV}/c, \nonumber\\
   && p_{\rm lab}^{\rm th}(\Xi^0p\rightarrow\Sigma^0\Sigma^+)
                  = 968.2 \mbox{ MeV}/c.
\end{eqnarray}
For $(\Lambda\Lambda,\Xi^0n,\Xi^-p,\Sigma^0\Lambda,\Sigma^0\Sigma^0,
\Sigma^-\Sigma^+)$:
\begin{eqnarray}
   && p_{\rm lab}^{\rm th}(\Lambda\Lambda\rightarrow\Xi^0n)
                  = 326.0 \mbox{ MeV}/c, \nonumber\\
   && p_{\rm lab}^{\rm th}(\Lambda\Lambda\rightarrow\Xi^-p)
                  = 361.2 \mbox{ MeV}/c, \nonumber\\
   && p_{\rm lab}^{\rm th}(\Lambda\Lambda\rightarrow\Sigma^0\Lambda)
                  = 611.3 \mbox{ MeV}/c, \nonumber\\
   && p_{\rm lab}^{\rm th}(\Lambda\Lambda\rightarrow\Sigma^0\Sigma^0)
                  = 900.8 \mbox{ MeV}/c, \nonumber\\
   && p_{\rm lab}^{\rm th}(\Lambda\Lambda\rightarrow\Sigma^-\Sigma^+)
                  = 906.6 \mbox{ MeV}/c.
\end{eqnarray}
For $(\Xi^-n,\Sigma^-\Lambda,\Sigma^-\Sigma^0)$:
\begin{eqnarray}
   && p_{\rm lab}^{\rm th}(\Xi^-n\rightarrow\Sigma^-\Lambda)
                  = 593.1 \mbox{ MeV}/c, \nonumber\\
   && p_{\rm lab}^{\rm th}(\Xi^-n\rightarrow\Sigma^-\Sigma^0)
                  = 972.8 \mbox{ MeV}/c.
\end{eqnarray}
For $(\Xi^-\Lambda,\Xi^0\Sigma^-,\Xi^-\Sigma^0)$:
\begin{eqnarray}
   && p_{\rm lab}^{\rm th}(\Xi^-\Lambda\rightarrow\Xi^0\Sigma^-)
                  = 685.4 \mbox{ MeV}/c, \nonumber\\
   && p_{\rm lab}^{\rm th}(\Xi^-\Lambda\rightarrow\Xi^-\Sigma^0)
                  = 692.9 \mbox{ MeV}/c.
\end{eqnarray}
For $(\Xi^0\Lambda,\Xi^0\Sigma^0,\Xi^-\Sigma^+)$:
\begin{eqnarray}
   && p_{\rm lab}^{\rm th}(\Xi^0\Lambda\rightarrow\Xi^0\Sigma^0)
                  = 690.4 \mbox{ MeV}/c, \nonumber\\
   && p_{\rm lab}^{\rm th}(\Xi^0\Lambda\rightarrow\Xi^-\Sigma^+)
                  = 705.9 \mbox{ MeV}/c.
\end{eqnarray}
If one uses the nonrelativistic approximation to relate the laboratory
momentum and total energy, the threshold momenta are found to be
lower, where the difference can be as large as 75 MeV/$c$.

\subsection{Meson-mass corrections}
Following the scheme of Refs.~\cite{Tho70,Rij91,Rij92,Rij96}, we start
out from the four-dimensional one-meson-exchange Feynman diagram and
end up with two three-dimensional time-ordered diagrams. The propagator
(energy denominator) for these two diagrams reads
\begin{equation}
   D(\omega)=\frac{1}{2\omega}\left[\frac{1}{E_2+E_3-W+\omega}
             +\frac{1}{E_1+E_4-W+\omega}\right].
\end{equation}
Here, $W=\sqrt{s}$ is the total energy and $\omega^2={\bf k}^2+m^2$,
with $m$ the meson mass and ${\bf k}={\bf p}'-{\bf p}$ the momentum
transfer. In the static approximation $E_i\rightarrow M_i$ and
$W\rightarrow M^0_1+M^0_2$, where we have included a superscript 0
to indicate that these masses refer to the masses of the particular
interaction channel we are considering. They are not necessarily
equal to the masses $M_1$ and $M_2$ occurring in the time-ordered
diagrams. For example, the potential for the $\Sigma\Sigma$
contribution in the coupled-channel $\Lambda\Lambda$ system has
$M_1=M_2=M_{\Sigma}$, but $M^0_1=M^0_2=M_{\Lambda}$.
In principle, the propagator in the static approximation can be
handled exactly using the fact that~\cite{Rij92}
\begin{equation}
  \frac{1}{\omega(\omega+a)}=\frac{2}{\pi}\int_0^{\infty}
          \frac{ad\lambda}{(\omega^{2}+\lambda^2)(a^2+\lambda^2)}
          +\frac{2\theta(-a)}{\omega^2-a^2},\ \ \ (a<m). \label{oma}
\end{equation}
This integral needs to be evaluated numerically whenever $a\neq0$,
which can be a considerable time factor in practical calculations.
We therefore make the same approximation as in Ref.~\cite{Rij99} and
use the fact that $M^0_1+M^0_2$ in most cases is rather close to half
the sum of initial- and final-state baryon masses.
The advantage of this, more crude, approximation is that the propagator
can then be written as
\begin{equation}
   D(\omega)\rightarrow \frac{1}{\omega^2-{\textstyle\frac{1}{4}}
            (M_3-M_4+M_2-M_1)^2},       \label{Deffmass}
\end{equation}
which means we have introduced an effective meson mass $\overline{m}$,
where the mass has dropped to
\begin{equation}
    m^2 \rightarrow \overline{m}^2=m^2-{\textstyle\frac{1}{4}}
                 (M_3-M_4+M_2-M_1)^2.
\end{equation}
In our potentials we have included the decrease in the physical pion
mass of 138.041 MeV/$c^2$ to 132.58 MeV/$c^2$ in
$\Lambda B\rightarrow\Sigma B$, where $B$ can be $N$, $\Sigma$, or
$\Xi$, and the much more significant decrease to 114.62 MeV/$c^2$ in
$\Sigma\Lambda\rightarrow\Lambda\Sigma$; in all other cases, we retain
the physical value of 138.041 MeV/$c^2$.
The $K$ and $K^*$ masses need to be reduced in all cases, where the
drop in mass squared ranges from $(125.56\mbox{ MeV}/c^2)^2$ in
$\Xi\Sigma\rightarrow\Sigma\Xi$ to $(253.63\mbox{ MeV}/c^2)^2$ in
$\Sigma N\rightarrow N\Sigma$.
Again, we argue that the scarce scattering data for the $S=-1$
sector and the absence of any scattering data for the $S<-1$ sectors
at this point in time does not yet warrant the more sophisticated
treatment of using Eq.~(\ref{oma}).

\section{Results}
\label{sec:results}
The main purpose of this paper is to present the properties of the
six NSC97 potentials for the $S=-2$, $-3$, and $-4$ sectors.
The free parameters in each model are fitted to the $N\!N$ and $Y\!N$
scattering data for the $S=0$ and $S=-1$ sectors, respectively.
Given the expressions for the coupling constants in terms of the octet
and singlet parameters, and their values for the six different models
as presented in Ref.~\cite{Rij99}, it is straightforward to evaluate
all possible baryon-baryon-meson coupling constants needed for the
$S\leq-2$ potentials. A complete set of coupling constants for models
NSC97a and NSC97f is given in Tables~\ref{tabcopa} and \ref{tabcopf},
respectively. In almost all cases, the coupling constants for the other
models, NSC97b--e, smoothly interpolate between these two extremes.

In the following we will present the model predictions for scattering
lengths, bound states, and cross sections. In order to present a
complete set of results for all the allowed baryon-baryon channels,
we will also include the results for the $S=0$ and $S=-1$ sectors.
Additional results and applications for the $Y\!N$ models in the
$S=-1$ sector can be found in Ref.~\cite{Rij99}.

\subsection{Effective-range parameters}
In Tables~\ref{tabscat0} through \ref{tabscat-4} we give the scattering
lengths and effective ranges for the singlet $^1S_0$ and triplet $^3S_1$
partial waves. The Coulomb interaction is included whenever the two
scattering baryons are charged. We observe the general trend that for
a particular value of the total strangeness, the change in scattering
length from one model to the next in most cases nicely follows the
change in $\alpha_V^m$ (which, in principle, labels each NSC97 model).

The $S=0$ results of Table~\ref{tabscat0} are included so as to present
the complete set of allowed channels. Here we also included the
experimental values as found in Refs.~\cite{Ber88,Koe75,Ter87}.
We clearly see that the present $N\!N$ models only encompass the
qualitative features of the $N\!N$ interaction. It is well-known that
the differences between the experimental singlet scattering lengths
and effective ranges for $pp$, $np$, and $nn$ cannot be explained
within a simple one-boson-exchange model (one needs additional
contributions from meson-mixing, two-pion exchange, pion-photon
exchange, etc.) and, indeed, for the NSC97 models presented here
we also find that the $np$ $a_s$ and $r_s$ are almost the same as
the $nn$ ones. The difference with the $pp$ ones are due to the
inclusion of the Coulomb interaction, of course.
However, these differences between the experimental values are only of
importance at low energies ($T_{\rm lab}\lesssim5$ MeV), where the
accuracy of the experimental $pp$ and $np$ scattering data magnifies
this charge dependence~\cite{Sto93}, and so the description of the
scattering data above $\sim5$ MeV can still be very acceptable.

In Table~\ref{tabscat-1} we repeat the $S=-1$ results for the scattering
lengths from our previous publication~\cite{Rij99}, but here we also
include the effective ranges. We note that also the effective ranges
for the different models exhibit a very similar trend with the value
of $\alpha_V^m$ as do the scattering lengths.

For the $S=-2$ sector the results are given in Table~\ref{tabscat-2}.
The $\Lambda\Lambda(^1S_0)$ scattering lengths are found to be rather
small, indicating a mildly attractive $\Lambda\Lambda$ interaction.
Experimental information on the ground states of
$_{\Lambda\Lambda}^{\,\ 6}$He, $_{\Lambda\Lambda}^{\;10}$Be, and
$_{\Lambda\Lambda}^{\;13}$B~\cite{Dal89}, seems to indicate a
separation energy of $\Delta B_{\Lambda\Lambda}=4-5$ MeV,
corresponding to a rather strong attractive $\Lambda\Lambda$
interaction. As a matter of fact, an estimate for the $\Lambda\Lambda$
$^1S_0$ scattering length, based on such a value for $\Delta
B_{\Lambda\Lambda}$, gives $a_{\Lambda\Lambda}(^1S_0)\approx-2.0$
fm~\cite{Tan65,Bod65}.
In the present approach, we can only increase the attraction in
the $\Lambda\Lambda$ channel by modifying the scalar-exchange
potential. However, if the scalar mesons are viewed as being mainly 
$q\bar{q}$ states, one finds that the (attractive) scalar-exchange
part of the interaction in the various channels satisfies
\begin{equation}
   |V_{\Lambda\Lambda}| < |V_{\Lambda N}| < |V_{N\!N}|.
\end{equation}
The NSC97 fits to the $Y\!N$ scattering data~\cite{Rij99} give values
for the scalar-meson mixing angle which seem to point to almost ideal
mixing for the scalars as $q\bar{q}$ states, and so an increased
attraction in the $\Lambda\Lambda$ channel would give rise to
(experimentally unobserved) bound states in the $\Lambda N$ channel.
On the other hand, preliminary results from a potential model which also
includes two-meson exchanges within the present framework~\cite{Rij98},
do show the apparently required attraction in the $\Lambda\Lambda$
interaction without giving rise to $\Lambda N$ bound states.

The large values for the triplet effective range $r_t$ in $\Xi^0p$
and $\Xi^-n$ are a simple reflection of the fact that the $^3S_1$
phase shift at small laboratory momenta is very small and only very
slowly increases in magnitude. The $^3S_1$ phase shift for models
NSC97a, NSC97b, and NSC97f starts off positive (hence, negative $a_t$),
does not exceed $0.3^\circ$ at $p_{\rm lab}=100$ MeV/$c$, and then
becomes negative at about 175 MeV/$c$. The $^3S_1$ phase shift for
models NSC97c--e starts off negative, but is still only about
$-1.0^\circ$ at $p_{\rm lab}=125$ MeV/$c$.

The sizable positive scattering lengths for $\Sigma^+\Sigma^+$
and $\Sigma^-\Sigma^-$ are a signal for bound states, which will
be discussed in the next subsection.

The effective-range parameters for the $S=-3$ sector are given in
Table~\ref{tabscat-3}. In this case, the large values for the singlet
and triplet effective ranges in $\Xi^-\Sigma^-$ are a reflection of the
fact that the presence of the Coulomb interaction causes the $^1S_0$
and $^3S_1$ phase shifts to start off very flatly at 180$^\circ$.
Removing the Coulomb interaction also removes the extreme flatness,
as can be seen from the much more modest values for the effective
ranges in $\Xi^0\Sigma^+$.

Finally, the effective-range parameters for the $S=-4$ sector are given
in Table~\ref{tabscat-4}. Also in this case, the positive values for
the singlet scattering lengths indicate bound states; see the next
subsection.

\subsection{Bound states in $\protect\bbox{S}$ waves}
Because the $N\!N$ triplet scattering length is slightly off (see
Table~\ref{tabscat0}), it is not surprising that the NSC97 results for
the deuteron are slightly off as well. The binding energies range from
--2.19 MeV for NSC97a to --2.07 MeV for NSC97f, which is to be compared
to the experimental deuteron binding energy of --2.224\,575(9)
MeV~\cite{Leu82}. In view of these results, we want to stress once
more that the NSC97 $N\!N$ potentials are here only included for
reasons of completeness. They should not be used for any high-precision
quantitative calculations, for which much more suitable potential
models can be found in the literature~\cite{Sto94,Wir95,Mac96}.

It turns out that the presence or absence of bound states in the other
interaction channels can best be understood in terms of the SU(3)
irreducible representations (irreps) to which the channel belongs,
and the well-known fact that the $N\!N$ interaction has a bound
state in the $^3S_1$-$^3D_1$ coupled partial wave (the deuteron)
and a quasi bound state in the $^1S_0$ partial wave. If an
interaction belongs to the same irrep as the $N\!N$ interaction,
unbroken SU(3) symmetry would imply that it also exhibits a
bound state or quasi bound state. However, SU(3) is not an exact
symmetry: the nucleons, hyperons, and cascades have different masses.
But it is possible that remnants of these (quasi) bound states can
still appear or that quasi bound states turn into truly bound states.
In order to make this comparison, we list in Table~\ref{tabirrep} all
the irreps to which the various baryon-baryon interactions belong, as
derived from details given in Ref.~\cite{Swa63}.

The $N\!N$, $\Sigma N$, $\Sigma\Sigma$, $\Xi\Sigma$ and $\Xi\Xi$
$^1S_0$ interactions all belong to the same $\{27\}$ irrep. For
these interactions this is also the only irrep. The $N\!N(^1S_0)$
interaction has a quasi bound state, and so we also expect (quasi)
bound states in the other channels. This is indeed what we find.
The effective ``potential'' $W$ for these interactions is shown in
Fig.~\ref{figW27}, where we refer to Ref.~\cite{Mae89} for the
definition of $W$. We only show the results for models NSC97a and
NSC97f as an example; the other models show very similar behavior.
We note that the short-range repulsion increases with the reduced
mass of the system, except for the $\Xi\Xi$ interaction which has
less repulsion than the $\Xi\Sigma$ interaction. The attractive
tails of the $N\!N$ and $\Sigma N$ interactions are almost identical,
and so it is not surprising that we also find a quasi bound state
in $\Sigma^+p$ and $\Sigma^-n$; note that the scattering lengths 
are rather similar to those of $pp$ and $nn$.

The tail of the $\Sigma\Sigma$ interaction is almost twice as strong;
strong enough to support a bound state. The presence of bound states
could already be inferred from the relatively large positive scattering
lengths for these systems; see Table~\ref{tabscat-2}.
The binding energies in $\Sigma^+\Sigma^+$ range from --1.53 MeV for
NSC97a to --3.07 MeV for NSC97f, while in $\Sigma^-\Sigma^-$ they
range from --1.59 MeV for NSC97a to --3.17 MeV for NSC97f.

The attraction in $\Xi\Sigma$ is even stronger and extends to smaller
inter-baryon distances. The binding energies in $\Xi^0\Sigma^+$
range from --3.02 MeV for NSC97a to --16.5 MeV for NSC97f. The
presence of the Coulomb interaction in $\Xi^-\Sigma^-$ causes a
shift of roughly 1 MeV, resulting in binding energies of --2.30 MeV
for NSC97a to --15.6 MeV for NSC97f.

Finally, the attraction in $\Xi\Xi$ is also strong enough to support
a bound state. The $\Xi^0\Xi^0$ and $\Xi^-\Xi^0$ give almost
identical results ranging from --0.10 MeV for NSC97a to --15.8 MeV
for NSC97f. Again, the presence of the Coulomb interaction in
$\Xi^-\Xi^-$ causes a shift of about 1 MeV, and so the NSC97a model
no longer supports a bound state in this channel.

The $N\!N$ and $\Xi\Sigma$ $^3S_1$-$^3D_1$ interactions both belong
to the $\{10^*\}$ irrep, and so, in analogy with the deuteron bound
state in $N\!N$, we also expect bound states in $\Xi\Sigma$. Indeed,
in $\Xi^0\Sigma^+$ the binding energies range from --5.64 MeV for
NSC97a to --36.1 MeV for NSC97f, whereas in $\Xi^-\Sigma^-$ they range
from --4.86 MeV for NSC97a to --35.3 MeV for NSC97f.

The fact that the nucleons and cascades both form isodoublets might
suggest that the $N\!N$ and $\Xi\Xi$ interactions are very similar.
Although this is true for the $I=1$ partial waves (both interactions
belong to the $\{27\}$ irrep), this is not the case for the $I=0$
partial waves. In Table~\ref{tabirrep} we see that the $I=0$ $N\!N$
interaction belongs to the $\{10^*\}$ irrep, while the $I=0$
$\Xi\Xi$ interaction belongs to the $\{10\}$ irrep. The $\{10^*\}$
supports a bound state (the deuteron), but apparently the $\{10\}$
does not, as can be deduced from the fact that a bound state in
the $^3S_1$-$^3D_1$ $\Sigma^+p$ or $\Sigma^-n$ channel has never
been found. This explains why here we also do not find a bound state
in $\Xi^-\Xi^0$: there is no deuteron-analogue in $\Xi\Xi$.

An analysis on the presence or absence of bound states in all the
other interaction channels is much more difficult. The reason, of
course, is that none of these remaining interactions belongs to one
single irrep. Hence, it is possible that the presence of one particular
irrep might provide enough attraction to support a bound state,
but that the presence of another irrep reduces this attraction
and prevents the existence of a bound state. The analysis is further
complicated by the fact that these other interactions (on the particle
basis) do not belong to pure isospin states, and so there is an
additional mixing of contributions form different irreps.
Therefore, we here do not attempt to analyze these channels and simply
suffice by stating that we do not find $S$-wave bound states in any
of them.

\subsection{Total cross sections}
We next present the predictions for the total cross section for all
$(S,Q)$ channels. Whenever a coupling to different channels is
involved (i.e., for the $\Lambda N$, $\Lambda\Lambda$, $\Xi N$, and
$\Xi\Lambda$ interactions), we only show the result for the leading
channel. For those cases where both baryons are charged, we do not
include the purely Coulomb contribution to the total cross section,
but we do include the Coulomb interference to the nuclear amplitude.
The cross section is calculated by summing the contributions from
partial waves with orbital angular momentum up to and including $L=2$.
We find this to be sufficient for all the $S\neq0$ sectors; inclusion
of any higher partial waves has no significant effect. In the case of
$N\!N$ scattering ($S=0$) we go up to $L=4$, which is high enough to
capture the general energy dependence of the $N\!N$ total cross section.
Inclusion of higher partial waves will shift the total cross section
to slightly higher values without changing the overall shape. Of course,
their inclusion would be necessary if a detailed comparison with real
experimental data were to be made.

In Fig.~\ref{figNNsgt}(a) we give the $nn$, $np$, and purely nuclear
$pp$ total cross sections for model NSC97a. The curves for the other
five models NSC97b--f are indistinguishable on this scale, and are
therefore left out. This similarity is a reflection of the fact that
all six models give an equally good description of the $S=0$ sector.
The purely nuclear $pp$ and $nn$ results are also almost identical which
is a consequence of the fact that the NSC97 $N\!N$ models do not contain
any explicit charge-symmetry breaking, and so any difference is totally
due to the neutron-proton mass difference, which is very small.
All three cross sections look very similar, but at $T_{\rm lab}=300$ MeV
the $np$ cross section of $\sim$39 mb is slightly lower than the $nn$
and $pp$ cross sections of $\sim$51 mb. Below $T_{\rm lab}\lesssim200$ MeV
the $np$ cross section becomes larger, while at very small laboratory
kinetic energies (not visible in this figure) the $nn$ and $pp$ cross
sections rapidly exceed the $np$ cross section again by more than a
factor of two. 
In Fig.~\ref{figNNsgt}(b) we also give the $pp$ result where we include
the modification due to the Coulomb-nuclear interference. In order to
illustrate the fact that this interference cross section vanishes as
the laboratory kinetic energy approaches zero, we have included a
subplot where we have expanded the 0--1 MeV region.

In Fig.~\ref{figYNsgt} we give the $\Sigma^-n$, $\Sigma^+p$, $\Lambda p$,
and $\Lambda n$ total cross sections. The purely nuclear $\Sigma^+p$
total cross section is very similar to the $\Sigma^-n$ one, and so here
we only give the $\Sigma^+p$ cross section including the Coulomb-nuclear
interference modification. Note that for the $S=-1$ sector (and all
other $S\neq0$ sectors to be presented below), the total cross section
is given as a function of laboratory momentum $p_{\rm lab}$, rather
than of laboratory kinetic energy $T_{\rm lab}$. The reason is that
experimental data for $N\!N$ scattering are usually given at a certain
$T_{\rm lab}$, whereas for $Y\!N$ scattering they are usually given at
a certain $p_{\rm lab}$. The $\Lambda p$ and $\Lambda n$ total cross
sections show a cusp effect when the $\Lambda p\rightarrow\Sigma^+n$
and $\Lambda n\rightarrow\Sigma^0n$ thresholds open up. The cusp is
due to the enhancement in the $^3S_1$ waves, which is caused by the
coupling of the $\Lambda N$ and $\Sigma N$ channels and the rather
strong interaction in the $^3S_1$-wave $\Sigma N$ channel.
For $\Lambda n$ we also observe a small bump when the threshold to
$\Sigma^-p$ opens up.
Again we note that the curves for the six NSC97 models are very close
to each other, which reflects the fact that these models all describe
the (scarce) $Y\!N$ scattering data equally well~\cite{Rij99}.
The spread in the curves at low momenta corresponds to a similar spread
in the scattering lengths; see Table~\ref{tabscat-1}.

In Fig.~\ref{figXNsgt} we present the $\Lambda\Lambda$, $\Xi^-n$ and
$\Xi^0p$, and $\Sigma^+\Sigma^+$ total cross sections; the latter for
both the purely nuclear case and the case including the Coulomb-nuclear
interference modification. The two corresponding $\Sigma^-\Sigma^-$
total cross sections are left out, since they are almost exactly the
same as the $\Sigma^+\Sigma^+$ ones.
Here, for the first time, the differences between the six models
clearly manifest themselves. The value for the $\Lambda\Lambda$ total
cross section at small momenta varies by almost a factor of four,
while at high momenta there is also a variation of at least a factor
of two. It is interesting that only NSC97e exhibits a smoothed-out
cusp effect when the two $\Lambda\Lambda\rightarrow\Xi N$ thresholds
open up, whereas the other models exhibit no such enhancement.
The six NSC97 results for the two $\Xi N$ total cross sections are
all very similar up to laboratory momenta close to 590 MeV/$c$, where
the thresholds to $\Sigma^-\Lambda$ or $\Sigma^+\Lambda$ open up. All
models exhibit a clear cusp effect, NSC97f being the most pronounced.
We also observe a cusp effect due to the opening up of the
$\Sigma^-\Sigma^0$ and $\Sigma^0\Sigma^+$ channels, but for models
NSC97e and NSC97f this effect is shifted to laboratory momenta lower than
the actual threshold value (about 970 MeV/$c$). A possible explanation
for this phenomenon is that the transition potential has enough
attraction to cause a virtual bound state, which manifests itself
as a multichannel resonance, rather than a cusp.
The NSC97 results for the $\Sigma\Sigma$ total cross sections show some
variation at low momenta, while for $p_{\rm lab}\gtrsim200$ MeV/$c$ all
six models give very similar results. The variation at low momenta
could already be inferred from inspecting the $\Sigma\Sigma$ $^1S_0$
scattering lengths in Table~\ref{tabscat-2}: here the models show a
substantial variation as well.

The total cross sections for $\Xi^0\Sigma^+$ and $\Xi^-\Sigma^-$, and
for $\Xi^-\Lambda$ and $\Xi^0\Lambda$ are shown in Fig.~\ref{figXYsgt}.
The $\Xi\Sigma$ and $\Xi\Lambda$ total cross sections are found to be
rather similar to the $\Sigma N$ and $\Lambda N$ total cross sections,
at least as far as the general energy dependence is concerned.
It is also interesting to note that for high laboratory momenta the
$\Xi\Lambda$ total cross sections are roughly of the same magnitude
as the $\Lambda N$ ones. These similarities can be understood from the
fact that both nucleons and cascades form isospin doublets, and so the
$S=-1$ and $S=-3$ interactions belong to the same set of SU(3) irreps;
see Table~\ref{tabirrep}. The only difference is an interchange
of $\{10\}$ and $\{10^*\}$. This might also be the reason why there is
no cusp effect in the $\Xi\Lambda$ cross section (at the threshold
momentum of 690 MeV/$c$): the isospin-1/2 ($\Lambda N,\Sigma N$)
interaction involves the $\{10^*\}$ irrep, to which also the deuteron
belongs, whereas the isospin-1/2 ($\Xi\Lambda,\Xi\Sigma$) interaction
involves the $\{10\}$ irrep. The small bump in the NSC97c cross section
is then probably due to the fact that the coupling to the isospin-3/2
$\Xi\Sigma$ interaction (which belongs to the $\{10^*\}$ irrep) in
this case is strong enough to cause an enhancement.
A further difference is that the $\Xi\Lambda$ and $\Xi\Sigma$ results
exhibit much more variation from one model to the next as do the
$\Lambda N$ and $\Sigma N$ results.

Finally, the total cross sections for $\Xi^0\Xi^0$, $\Xi^-\Xi^0$,
and $\Xi^-\Xi^-$ are given in Fig.~\ref{figXXsgt}. For $\Xi^-\Xi^-$
we give both the purely nuclear cross section (c) and the one
including the Coulomb-nuclear interference modification (d).
As expected, the results are similar to the $N\!N$ results, but again
the differences between the six NSC97 models are much more pronounced.
The NSC97a result for the total cross sections is found to be very
large, as was to be expected in view of the large $^1S_0$ scattering
length, given in Table~\ref{tabscat-4}.

\section{SUMMARY AND CONCLUSION}
\label{sec:conc}
As already stated in our previous publication~\cite{Rij99}, the NSC97
potentials presented here are an important step forward in modeling
the baryon-baryon interactions for scattering and hypernuclei in the
context of broken SU(3)$_F$ symmetry. The potentials are based on
the one-boson-exchange model, where the coupling constants at the
baryon-baryon-meson vertices are restricted by the broken SU(3) symmetry.
Each type of meson exchange (pseudoscalar, vector, scalar)
contains five free parameters: a singlet coupling constant, an octet
coupling constant, the $F/(F+D)$ ratio $\alpha$, a meson-mixing angle,
and a parameter $\lambda$ which effectively accounts for the fact that
the strange quark is much heavier than the up and down quarks.
However, they are not all treated as free parameters: the pseudoscalar
and vector $F/(F+D)$ parameters and meson-mixing angles are fixed
from other sources~\cite{Rij99}.
The potentials are regularized with exponential cutoff parameters,
which provide a few additional free parameters.
Most of these parameters are fixed in the fit to the wealth of
accurate $N\!N$ scattering data, while the remaining ones are fixed
in the fit to the (few) $Y\!N$ scattering data. Here we note that,
although the scattering data for the $Y\!N$ sector are very scarce,
they are extremely valuable in constraining $Y\!N$ potential models.
As a matter of fact, it is not at all trivial to obtain a good fit
to the $Y\!N$ data and at the same time avoid (experimentally
unobserved) bound states in the $\Lambda N$ and $\Sigma N$ channels.
However, there is still enough freedom to construct six different
models, NSC97a through NSC97f.
They all describe the $N\!N$ and $Y\!N$ data equally well, but differ
on a more detailed level. The assumption of SU(3) symmetry then
allows us to extend these models to the higher strangeness channels
(i.e., $YY$ and all interactions involving cascades), without the
need to introduce additional free parameters. The NSC97 models are
the first potential models of this kind.

In order to illustrate the basic properties of these potentials,
we have presented results for scattering lengths, possible bound
states in $S$-waves, and total cross sections. The results for the
six different models are rather similar for the $S=0$ ($N\!N$) and
$S=-1$ ($Y\!N$) sectors, but then each model was fitted such as to
ensure equally good descriptions of the data.
The predictions for the $S\leq-2$ sectors can be viewed as extrapolations
and, indeed, those results show much more variation from one model to
the next. It would, therefore, be very worthwhile to have experimental
information on the interactions in these $S\leq-2$ sectors to further
constrain the potentials, and to test the SU(3)-symmetry assumptions
that have been made. Information from hypernuclear data assumes
knowledge of how to treat many-body effects, and so two-body
scattering data would be preferred. However, we realize, of course,
that hyperons and cascades are short-lived and hard to produce in
large quantities, which makes it very difficult indeed to set up a
good scattering experiment.

Armed with the fact that we know that the $N\!N$ $^1S_0$ interaction,
belonging to the $\{27\}$ irrep, is attractive enough to form a
quasi bound state, and that the $N\!N$ $^3S_1$-$^3D_1$ interaction,
belonging to the $\{10^*\}$ irrep, causes the deuteron bound state,
we can understand the presence of bound states in some of the other
channels by virtue of the fact whether the corresponding interaction
involves the $\{27\}$ or $\{10^*\}$ irrep, or not. This is an interesting
observation, because our potentials do not obey a perfect SU(3)
symmetry. First of all, we use the physical baryon masses which are
substantially different from the SU(3) average. Second, the cutoff
parameters which regularize the vertices are not dictated by to what
irrep a particular interaction belongs~\cite{Mae89}, but rather by what
meson is being exchanged; for more details, see Ref.~\cite{Rij99}.
Finally, for each type of meson exchange, we have introduced a
parameter $\lambda$ which explicitly breaks the SU(3) symmetry
(in some models up to 20\%); see above and Ref.~\cite{Rij99}.
In spite of these modifications, our results show that the general
features of an exact SU(3) symmetry survive to a remarkable degree.
Again, experimental information on the $S\leq-2$ systems will be
invaluable as a test of these results.

We conclude by mentioning that these NSC97 potentials provide an
excellent starting point for calculations on multi-strange systems.
Unlike other approaches, these are the first models for which the
$S\leq-2$ interactions contain no free parameters, and for which the
$S=0$ and $S=-1$ interactions are fitted to the two-body scattering
data. They can be used to calculate properties of hypernuclei
(including double-$\Lambda$ or even more exotic hypernuclei) and
to explore strange nuclear matter. Our initial efforts for the
latter will be published elsewhere~\cite{Sto99}.

\acknowledgments
The authors would like to thank J.J.\ de Swart and T.-S.H.\ Lee for
many stimulating discussions. We would also especially like to thank
P.M.M.\ Maessen who did pioneering work on constructing a soft-core
model for the $S=-2$ interaction, which formed the starting point
for developing the present $S\leq-2$ NSC97 models.
The work of V.G.J.S.\ was partly supported by the U.S.\ Department of
Energy, Nuclear Physics Division, under Contract No.\ W-31-109-ENG-38.

\begin{table}
\caption{Baryon masses in MeV/$c^2$.}
\begin{tabular}{lcd}
  Baryon & & Mass \\
\tableline
  Nucleon & $p$            &  938.27231 \\
          & $n$            &  939.56563 \\
  Hyperon & $\Lambda$      & 1115.684   \\
          & $\Sigma^+$     & 1189.37    \\
          & $\Sigma^0$     & 1192.55    \\
          & $\Sigma^-$     & 1197.436   \\
  Cascade & $\Xi^0$        & 1314.90    \\
          & $\Xi^-$        & 1321.32
\end{tabular}
\label{tabmass}
\end{table}

\mediumtext
\begin{table}
\caption{Possible interaction channels on the isospin basis,
         labeled according to the total strangeness $S$ and
         total isospin $I$.}
\begin{tabular}{lccccc}
         & $I=0$ & $I={\textstyle\frac{1}{2}}$ & $I=1$
                 & $I={\textstyle\frac{3}{2}}$ & $I=2$ \\[1mm]
\tableline
  $S= 0$ & $N\!N$ & & $N\!N$ & & \\[1mm]
  $S=-1$ & & $(\Lambda N,\Sigma N)$ & & $\Sigma N$ & \\[1mm]
  $S=-2$ & $(\Lambda\Lambda,\Xi N,\Sigma\Sigma)$ &
         & $(\Xi N,\Sigma\Lambda,\Sigma\Sigma)$ & & $\Sigma\Sigma$ \\[1mm]
  $S=-3$ & & $(\Xi\Lambda,\Xi\Sigma)$ & & $\Xi\Sigma$ & \\[1mm]
  $S=-4$ & $\Xi\Xi$ & & $\Xi\Xi$ & &
\end{tabular}
\label{tabisobasis}
\end{table}

\widetext
\begin{table}
\caption{Possible interaction channels on the particle basis,
         labeled according to the total strangeness $S$ and
         total particle charge $Q$.}
\begin{tabular}{lccccc}
         & $Q=-2$ & $Q=-1$ & $Q=0$ & $Q=+1$ & $Q=+2$ \\
\tableline
  $S= 0$ & & & $nn$ & $np$ & $pp$ \\[1mm]
  $S=-1$ & & $\Sigma^-n$ & $(\Lambda n,\Sigma^0n,\Sigma^-p)$
         & $(\Lambda p,\Sigma^+n,\Sigma^0p)$ & $\Sigma^+p$ \\[1mm]
  $S=-2$ & $\Sigma^-\Sigma^-$ & $(\Xi^-n,\Sigma^-\Lambda,\Sigma^-\Sigma^0)$
         & $(\Lambda\Lambda,\Xi^0n,\Xi^-p,
             \Sigma^0\Lambda,\Sigma^0\Sigma^0,\Sigma^-\Sigma^+)$
         & $(\Xi^0p,\Sigma^+\Lambda,\Sigma^0\Sigma^+)$
         & $\Sigma^+\Sigma^+$ \\[1mm]
  $S=-3$ & $\Xi^-\Sigma^-$ & $(\Xi^-\Lambda,\Xi^0\Sigma^-,\Xi^-\Sigma^0)$
         & $(\Xi^0\Lambda,\Xi^0\Sigma^0,\Xi^-\Sigma^+)$
         & $\Xi^0\Sigma^+$ & \\[1mm]
  $S=-4$ & $\Xi^-\Xi^-$ & $\Xi^-\Xi^0$ & $\Xi^0\Xi^0$ & & 
\end{tabular}
\label{tabpartbasis}
\end{table}

\narrowtext
\begin{table}
\caption{Isospin factors for the various meson exchanges in the
         different total strangeness and isospin channels.
         $P$ is the exchange operator.
         The $I=2$ case only contributes to $S=-2$ $\Sigma\Sigma$
         scattering where the isospin factors can collectively be
         given by $(\Sigma\Sigma|\eta,\eta',\pi|\Sigma\Sigma)=
         {\protect\textstyle\frac{1}{2}}(1+P)$, and so they are
         not separately displayed in the table.}
\begin{tabular}{ccc}
  $S=0$ & $I=0$ & $I=1$ \\
\tableline
  $(N\!N|\eta,\eta'|N\!N)$ & ${\textstyle\frac{1}{2}}(1-P)$
                           & ${\textstyle\frac{1}{2}}(1+P)$ \\[.5mm]
  $(N\!N|\pi|N\!N)$        & $-{\textstyle\frac{3}{2}}(1-P)$
                           & ${\textstyle\frac{1}{2}}(1+P)$ \\[3mm]
  $S=-1$ & $I={\textstyle\frac{1}{2}}$
         & $I={\textstyle\frac{3}{2}}$ \\[1mm]
\tableline
  $(\Lambda N|\eta,\eta'|\Lambda N)$  &    1         &     0 \\
  $(\Sigma  N|\eta,\eta'|\Sigma  N)$  &    1         &     1 \\
  $(\Sigma  N|\pi|\Sigma N)$          &  $-2$        &     1 \\
  $(\Lambda N|\pi|\Sigma N)$          & $-\sqrt{3}$  &     0 \\
  $(\Lambda N|K|N \Lambda)$           &   $P$        &     0 \\
  $(\Sigma  N|K|N \Sigma)$            &  $-P$        &  $2P$ \\
  $(\Lambda N|K|N \Sigma)$            & $-P\sqrt{3}$ &     0 \\[3mm]
  $S=-2$ & $I=0$ & $I=1$ \\
\tableline
  $(\Lambda\Lambda|\eta,\eta'|\Lambda\Lambda)$
                       & ${\textstyle\frac{1}{2}}(1+P)$ & 0 \\[.5mm]
  $(\Xi N|\eta,\eta'|\Xi N)$ & ${\textstyle\frac{1}{2}}(1+P)$ & 1 \\[.5mm]
  $(\Sigma\Sigma|\eta,\eta'|\Sigma\Sigma)$
                       & ${\textstyle\frac{1}{2}}(1+P)$
                       & ${\textstyle\frac{1}{2}}(1-P)$ \\
  $(\Sigma\Lambda|\eta,\eta'|\Sigma\Lambda)$ & 0 & 1 \\
  $(\Xi N|\pi|\Xi N)$  & $-3$ & 1 \\
  $(\Sigma\Sigma|\pi|\Sigma\Sigma)$ & $-(1+P)$
                       & $-{\textstyle\frac{1}{2}}(1-P)$ \\[.5mm]
  $(\Lambda\Lambda|\pi|\Sigma\Sigma)$
                       & $-{\textstyle\frac{1}{2}}\sqrt{3}(1+P)$ & 0 \\
  $(\Sigma\Lambda|\pi|\Lambda\Sigma)$ & 0 & $P$ \\
  $(\Sigma\Sigma|\pi|\Sigma\Lambda)$ & 0 & $(1-P)$ \\
  $(\Lambda\Lambda|K|\Xi N)$ & $1+P$ & 0 \\
  $(\Sigma\Sigma|K|\Xi N)$   & $\sqrt{3}(1+P)$ & $\sqrt{2}(1-P)$ \\
  $(\Xi N|K|\Sigma\Lambda)$  & 0 & $\sqrt{2};-P\sqrt{2}$ \\[3mm]
  $S=-3$ & $I={\textstyle\frac{1}{2}}$
         & $I={\textstyle\frac{3}{2}}$ \\[1mm]
\tableline
  $(\Xi\Lambda|\eta,\eta'|\Xi\Lambda)$  &    1        &     0 \\
  $(\Xi\Sigma |\eta,\eta'|\Xi\Sigma)$   &    1        &     1 \\
  $(\Xi\Sigma |\pi|\Xi\Sigma)$          &  $-2$       &     1 \\
  $(\Xi\Lambda|\pi|\Xi\Sigma)$          & $\sqrt{3}$  &     0 \\
  $(\Xi\Lambda|K|\Lambda\Xi)$           &   $P$       &     0 \\
  $(\Xi\Sigma |K|\Sigma\Xi)$            &  $-P$       &  $2P$ \\
  $(\Xi\Lambda|K|\Sigma\Xi)$            & $P\sqrt{3}$ &     0 \\[3mm]
  $S=-4$ & $I=0$ & $I=1$ \\
\tableline
  $(\Xi\Xi|\eta,\eta'|\Xi\Xi)$ & ${\textstyle\frac{1}{2}}(1-P)$
                               & ${\textstyle\frac{1}{2}}(1+P)$ \\[.5mm]
  $(\Xi\Xi|\pi|\Xi\Xi)$        & $-{\textstyle\frac{3}{2}}(1-P)$
                               & ${\textstyle\frac{1}{2}}(1+P)$
\end{tabular}
\label{tabiso}
\end{table}

\begin{table}
\caption{Coupling constants for model NSC97a, divided by
         $\protect\sqrt{4\pi}$. $M$ refers to the meson.
         The coupling constants are listed in the order pseudoscalar,
         vector ($g$ and $f$), scalar, and diffractive.}
\begin{tabular}{ccrrrrccrrrr}
 Type & $M$ & $N\!N\!M$ & $\Sigma\Sigma M$ & $\Sigma\Lambda M$ & $\Xi\Xi M$
 & $\hspace*{2ex}$
 & $M$ & $\Lambda N\!M$ & $\Lambda\Xi M$ & $\Sigma N\!M$ & $\Sigma\Xi M$ \\
\tableline
 $f$ & $\pi$    &   0.2729 &   0.1937 &   0.2032 & --0.0791
    && $K$      & --0.2578 &   0.0633 &   0.0757 & --0.2612 \\
 $g$ & $\rho$   &   0.8369 &   1.6738 &   0.0000 &   0.8369
    && $K^*$    & --1.2009 &   1.2009 & --0.6933 & --0.6933 \\
 $f$ &          &   3.5317 &   3.1409 &   2.2647 & --0.3908
    &&          & --3.1917 &   1.3154 &   0.3238 & --2.9260 \\
 $g$ & $a_0$    &   1.3951 &   3.0301 & --0.1385 &   1.6350
    && $\kappa$ & --2.3448 &   2.4720 & --1.5006 & --1.2804 \\
 $g$ & $a_2$    &   0.0000 &   0.0000 &   0.0000 &   0.0000
    && $K^{**}$ &   0.0000 &   0.0000 &   0.0000 &   0.0000 \\[2mm]
 Type & $M$ & $N\!N\!M$ & $\Lambda\Lambda M$ & $\Sigma\Sigma M$ & $\Xi\Xi M$
 & $\hspace*{2ex}$
 & $M$ & $N\!N\!M$ & $\Lambda\Lambda M$ & $\Sigma\Sigma M$ & $\Xi\Xi M$ \\
\tableline
 $f$ & $\eta$        &   0.1331 & --0.1210 &   0.2560 & --0.1654
    && $\eta'$       &   0.1441 &   0.2351 &   0.0975 &   0.2483 \\
 $g$ & $\omega$      &   2.9213 &   1.3995 &   2.2521 &   0.5499
    && $\phi$        & --0.4145 & --1.0738 & --1.7281 & --1.7397 \\
 $f$ &               &   1.1834 & --1.2299 &   1.4740 & --1.5653
    &&               &   1.0933 & --1.9109 &   1.4251 & --2.3011 \\
 $g$ & $\varepsilon$ &   4.6564 &   2.6380 &   3.1131 &   1.0709
    && $f_0$         & --0.1423 & --1.8195 & --2.5165 & --3.5623 \\
 $g$ & $P$           &   2.2722 &   2.2722 &   2.2722 &   2.2722
    && $f_2$         & --1.7435 & --1.7435 & --1.7435 & --1.7435
\end{tabular}
\label{tabcopa}
\end{table}

\begin{table}
\caption{Coupling constants for model NSC97f, divided by
         $\protect\sqrt{4\pi}$. $M$ refers to the meson.
         The coupling constants are listed in the order pseudoscalar,
         vector ($g$ and $f$), scalar, and diffractive.}
\begin{tabular}{ccrrrrccrrrr}
 Type & $M$ & $N\!N\!M$ & $\Sigma\Sigma M$ & $\Sigma\Lambda M$ & $\Xi\Xi M$
 & $\hspace*{2ex}$
 & $M$ & $\Lambda N\!M$ & $\Lambda\Xi M$ & $\Sigma N\!M$ & $\Sigma\Xi M$ \\
\tableline
 $f$ & $\pi$    &   0.2729 &   0.1937 &   0.2032 & --0.0791
    && $K$      & --0.3347 &   0.0822 &   0.0983 & --0.3390 \\
 $g$ & $\rho$   &   0.8369 &   1.6738 &   0.0000 &   0.8369
    && $K^*$    & --1.7222 &   1.7222 & --0.9943 & --0.9943 \\
 $f$ &          &   3.5317 &   2.5758 &   2.5909 & --0.9559
    &&          & --4.1896 &   1.1112 &   1.1357 & --4.1961 \\
 $g$ & $a_0$    &   1.3951 &   3.1758 & --0.2226 &   1.7807
    && $\kappa$ & --2.8237 &   3.0619 & --1.9053 & --1.4928 \\
 $g$ & $a_2$    &   0.0000 &   0.0000 &   0.0000 &   0.0000
    && $K^{**}$ &   0.0000 &   0.0000 &   0.0000 &   0.0000 \\[2mm]
 Type & $M$ & $N\!N\!M$ & $\Lambda\Lambda M$ & $\Sigma\Sigma M$ & $\Xi\Xi M$
 & $\hspace*{2ex}$
 & $M$ & $N\!N\!M$ & $\Lambda\Lambda M$ & $\Sigma\Sigma M$ & $\Xi\Xi M$ \\
\tableline
 $f$ & $\eta$        &   0.1331 & --0.0757 &   0.2800 & --0.2433
    && $\eta'$       &   0.1441 &   0.3417 &   0.0454 &   0.4491 \\
 $g$ & $\omega$      &   2.9213 &   2.8782 &   1.7592 &   1.9526
    && $\phi$        & --0.4145 & --2.2085 & --1.3498 & --3.9939 \\
 $f$ &               &   1.1834 & --0.7105 &   2.3234 & --1.9811
    &&               &   0.2709 & --2.7207 &   1.9311 & --4.5508 \\
 $g$ & $\varepsilon$ &   4.6564 &   3.3827 &   2.5449 &   1.4766
    && $f_0$         & --0.1423 & --2.5141 & --2.4120 & --5.0924 \\
 $g$ & $P$           &   2.2722 &   2.2722 &   2.2722 &   2.2722
    && $f_2$         & --1.7435 & --1.7435 & --1.7435 & --1.7435
\end{tabular}
\label{tabcopf}
\end{table}

\begin{table}
\caption{Singlet ($^1S_0$) and triplet ($^3S_1$) scattering lengths,
         $a_{s,t}$, and effective ranges, $r_{s,t}$, in fm
         for the different models in the total strangeness $S=0$
         ($N\!N$) sector. The experimental values are from
         Refs.~\protect\cite{Ber88,Koe75,Ter87}.}
\begin{tabular}{ccccccccc}
  & \multicolumn{2}{c}{$pp(^1S_0)$} & \multicolumn{2}{c}{$np(^1S_0)$}
  & \multicolumn{2}{c}{$np(^3S_1)$} & \multicolumn{2}{c}{$nn(^1S_0)$}\\
 Model & $a_s$ & $r_s$ & $a_s$ & $r_s$ & $a_t$ & $r_t$ & $a_s$ & $r_s$ \\
\tableline
 a & --7.48 & 2.75  & --15.79 & 2.83  & 5.43 & 1.72  & --15.88 & 2.83 \\
 b & --7.36 & 2.77  & --15.21 & 2.84  & 5.48 & 1.73  & --15.29 & 2.84 \\
 c & --7.27 & 2.78  & --14.81 & 2.85  & 5.52 & 1.74  & --14.88 & 2.85 \\
 d & --7.22 & 2.78  & --14.56 & 2.86  & 5.55 & 1.74  & --14.63 & 2.86 \\
 e & --7.20 & 2.79  & --14.49 & 2.86  & 5.56 & 1.74  & --14.56 & 2.86 \\
 f & --7.19 & 2.79  & --14.45 & 2.86  & 5.57 & 1.74  & --14.52 & 2.86 \\
 Exp. & --7.8063(26) & 2.794(14) & --23.749(8) & 2.81(5) 
      &   5.424(3)   & 1.760(5)  & --18.5(4)   & 2.80(11) \\
\end{tabular}
\label{tabscat0}
\end{table}

\begin{table}
\caption{Singlet ($^1S_0$) and triplet ($^3S_1$) scattering lengths,
         $a_{s,t}$, and effective ranges, $r_{s,t}$, in fm
         for the different models in the total strangeness $S=-1$
         ($Y\!N$) sector.}
\begin{tabular}{ccccccccc}
  & \multicolumn{2}{c}{$\Sigma^+p$} & \multicolumn{2}{c}{$\Lambda p$}
  & \multicolumn{2}{c}{$\Lambda n$} & \multicolumn{2}{c}{$\Sigma^-n$}\\
 Model & $a_s$ & $a_t$ & $a_s$ & $a_t$ & $a_s$ & $a_t$ & $a_s$ & $a_t$ \\
\tableline
 a & --4.35 & --0.14 & --0.71 & --2.18 & --0.77 & --2.15 & --6.06 & --0.18 \\
 b & --4.32 & --0.17 & --0.90 & --2.13 & --0.97 & --2.09 & --6.06 & --0.18 \\
 c & --4.28 & --0.25 & --1.20 & --2.08 & --1.28 & --2.07 & --5.98 & --0.28 \\
 d & --4.23 & --0.29 & --1.71 & --1.95 & --1.82 & --1.94 & --5.89 & --0.33 \\
 e & --4.23 & --0.28 & --2.10 & --1.86 & --2.24 & --1.83 & --5.90 & --0.32 \\
 f & --4.35 & --0.25 & --2.51 & --1.75 & --2.68 & --1.67 & --6.16 & --0.29
                                                                    \\[1.5mm]
 Model & $r_s$ & $r_t$ & $r_s$ & $r_t$ & $r_s$ & $r_t$ & $r_s$ & $r_t$ \\
\tableline
 a &   3.16 &--59.48 &   5.86 &   2.76 &   6.09 &   2.71 &   3.27 &--40.27 \\
 b &   3.17 &--43.24 &   4.92 &   2.84 &   5.09 &   2.80 &   3.28 &--29.28 \\
 c &   3.18 &--20.26 &   4.11 &   2.92 &   4.22 &   2.86 &   3.28 &--13.79 \\
 d &   3.19 &--16.78 &   3.46 &   3.08 &   3.52 &   3.01 &   3.29 &--11.29 \\
 e &   3.18 &--19.63 &   3.19 &   3.19 &   3.24 &   3.14 &   3.28 &--13.01 \\
 f &   3.14 &--25.35 &   3.03 &   3.32 &   3.07 &   3.34 &   3.24 &--16.44
\end{tabular}
\label{tabscat-1}
\end{table}

\begin{table}
\caption{Singlet ($^1S_0$) and triplet ($^3S_1$) scattering lengths,
         $a_{s,t}$, and effective ranges, $r_{s,t}$, in fm
         for the different models in the total strangeness $S=-2$
         ($YY$ and $\Xi N$) sector.}
\begin{tabular}{cccccccc}
  & $\Sigma^+\Sigma^+$ & \multicolumn{2}{c}{$\Xi^0p$} & $\Lambda\Lambda$
  & \multicolumn{2}{c}{$\Xi^-n$} & $\Sigma^-\Sigma^-$ \\
 Model & $a_s$ & $a_s$ & $a_t$ & $a_s$ & $a_s$ & $a_t$ & $a_s$ \\
\tableline
 a & 10.32 & 0.46 & --0.038 & --0.27 &  0.46 & --0.039 & 10.06 \\
 b &  9.96 & 0.45 & --0.045 & --0.38 &  0.45 & --0.046 &  9.72 \\
 c &  9.69 & 0.43 &   0.001 & --0.53 &  0.43 &   0.001 &  9.46 \\
 d &  8.77 & 0.42 &   0.041 & --0.53 &  0.42 &   0.040 &  8.58 \\
 e &  8.10 & 0.41 &   0.050 & --0.50 &  0.41 &   0.050 &  7.94 \\
 f &  6.98 & 0.40 & --0.030 & --0.35 &  0.40 & --0.031 &  6.85 \\[1.5mm]
 Model & $r_s$ & $r_s$ & $r_t$ & $r_s$ & $r_s$ & $r_t$ & $r_s$ \\
\tableline
 a &  1.60 & --6.12 & 661     & 15.00 & --6.09 & 634     &  1.59 \\
 b &  1.59 & --6.41 & 497     & 10.24 & --6.37 & 479     &  1.58 \\
 c &  1.57 & --6.97 & $>10^5$ &  7.43 & --6.92 & $>10^5$ &  1.57 \\
 d &  1.54 & --7.57 & 533     &  8.24 & --7.51 & 546     &  1.53 \\
 e &  1.51 & --8.08 & 339     &  9.11 & --8.01 & 346     &  1.50 \\
 f &  1.46 & --8.94 & 912     & 14.68 & --8.88 & 870     &  1.46
\end{tabular}
\label{tabscat-2}
\end{table}

\begin{table}
\caption{Singlet ($^1S_0$) and triplet ($^3S_1$) scattering lengths,
         $a_{s,t}$, and effective ranges, $r_{s,t}$, in fm
         for the different models in the total strangeness $S=-3$
         ($\Xi Y$) sector.}
\begin{tabular}{ccccccccc}
  & \multicolumn{2}{c}{$\Xi^0\Sigma^+$}
  & \multicolumn{2}{c}{$\Xi^0\Lambda$}
  & \multicolumn{2}{c}{$\Xi^-\Lambda$}
  & \multicolumn{2}{c}{$\Xi^-\Sigma^-$}\\
 Model & $a_s$ & $a_t$ & $a_s$ & $a_t$ & $a_s$ & $a_t$ & $a_s$ & $a_t$ \\
\tableline
 a & 4.13 & 3.21 & --0.80 &   0.54 & --0.83 &   0.52 & 0.34 & 0.23 \\
 b & 3.59 & 2.88 & --1.14 &   2.15 & --1.18 &   1.55 & 0.27 & 0.20 \\
 c & 3.11 & 2.96 & --1.81 & --0.27 & --1.86 & --0.34 & 0.22 & 0.21 \\
 d & 2.67 & 2.56 & --2.47 &   0.06 & --2.56 &   0.04 & 0.18 & 0.17 \\
 e & 2.47 & 2.23 & --2.65 &   0.17 & --2.75 &   0.17 & 0.16 & 0.14 \\
 f & 2.32 & 1.71 & --2.11 &   0.33 & --2.19 &   0.33 & 0.15 & 0.10 \\[1.5mm]
 Model & $r_s$ & $r_t$ & $r_s$ & $r_t$ & $r_s$ & $r_t$ & $r_s$ & $r_t$ \\
\tableline
 a & 1.46 & 1.28 &  4.71 & --0.47 &  4.79 & --0.41 & 2225 & 3295 \\
 b & 1.41 & 1.24 &  3.80 & --1.32 &  3.84 & --1.20 & 2795 & 3850 \\
 c & 1.35 & 1.28 &  3.11 &   9.83 &  3.13 &   6.60 & 3465 & 3720 \\
 d & 1.27 & 1.21 &  2.88 & 180.9  &  2.89 & 272.1  & 4300 & 4530 \\
 e & 1.22 & 1.12 &  2.89 &  15.15 &  2.88 &  15.54 & 4770 & 5440 \\
 f & 1.17 & 0.96 &  3.21 &   2.79 &  3.21 &   2.66 & 5170 & 7580
\end{tabular}
\label{tabscat-3}
\end{table}

\begin{table}
\caption{Singlet ($^1S_0$) and triplet ($^3S_1$) scattering lengths,
         $a_{s,t}$, and effective ranges, $r_{s,t}$, in fm
         for the different models in the total strangeness $S=-4$
         ($\Xi\Xi$) sector.}
\begin{tabular}{ccccccccc}
  & \multicolumn{2}{c}{$\Xi^0\Xi^0(^1S_0)$}
  & \multicolumn{2}{c}{$\Xi^-\Xi^0(^1S_0)$}
  & \multicolumn{2}{c}{$\Xi^-\Xi^0(^3S_1)$}
  & \multicolumn{2}{c}{$\Xi^-\Xi^-(^1S_0)$}\\
 Model & $a_s$ & $r_s$ & $a_s$ & $r_s$ & $a_t$ & $r_t$ & $a_s$ & $r_s$ \\
\tableline
 a & 17.81 & 1.85 & 17.28 & 1.85 & 0.40 &  3.45 & --1.27 & --609.3 \\
 b &  5.60 & 1.62 &  5.56 & 1.62 & 0.36 &  4.64 &   0.62 &  1069 \\
 c &  3.41 & 1.44 &  3.40 & 1.44 & 0.27 &  9.89 &   0.26 &  2686 \\
 d &  2.66 & 1.33 &  2.66 & 1.33 & 0.28 & 10.18 &   0.18 &  3840 \\
 e &  2.46 & 1.30 &  2.45 & 1.30 & 0.33 &  6.88 &   0.16 &  4290 \\
 f &  2.38 & 1.29 &  2.38 & 1.29 & 0.48 &  2.80 &   0.16 &  4465
\end{tabular}
\label{tabscat-4}
\end{table}

\begin{table}
\caption{SU(3) content of the different interaction channels.
         $S$ is the total strangeness and $I$ is the isospin.
         The upper half refers to the space-spin symmetric
         states $^3S_1$, $^1P_1$, $^3D$, \ldots, while the
         lower half refers to the space-spin antisymmetric
         states $^1S_0$, $^3P$, $^1D_2$, \ldots}
\begin{tabular}{cccc}
  \multicolumn{4}{c}{Space-spin symmetric states}                       \\
 $S$ & $I$ & Channels                  & SU(3) irreps                   \\
\tableline
   0 &   0 & $N\!N$                    & $\{10^*\}$                     \\
 --1 & 1/2 & $\Lambda N$, $\Sigma N$   & $\{10^*\}$, $\{8\}_a$          \\
     & 3/2 & $\Sigma N$                & $\{10\}$                       \\
 --2 &   0 & $\Xi N$                   & $\{8\}_a$                      \\
     &   1 & $\Xi N$, $\Sigma\Sigma$   & $\{10\}$, $\{10^*\}$, $\{8\}_a$\\
     &     & $\Sigma\Lambda$           & $\{10\}$, $\{10^*\}$           \\
 --3 & 1/2 & $\Xi\Lambda$, $\Xi\Sigma$ & $\{10\}$, $\{8\}_a$            \\
     & 3/2 & $\Xi\Sigma$               & $\{10^*\}$                     \\
 --4 &   0 & $\Xi\Xi$                  & $\{10\}$                       \\[2mm]
  \multicolumn{4}{c}{Space-spin antisymmetric states}                   \\
 $S$ & $I$ & Channels                  & SU(3) irreps                   \\
\tableline
   0 &   1 & $N\!N$                    & $\{27\}$                       \\
 --1 & 1/2 & $\Lambda N$, $\Sigma N$   & $\{27\}$, $\{8\}_s$            \\
     & 3/2 & $\Sigma N$                & $\{27\}$                       \\
 --2 &   0 & $\Lambda\Lambda$, $\Xi N$, $\Sigma\Sigma$
                                       & $\{27\}$, $\{8\}_s$, $\{1\}$   \\
     &   1 & $\Xi N$, $\Sigma\Lambda$  & $\{27\}$, $\{8\}_s$            \\
     &   2 & $\Sigma\Sigma$            & $\{27\}$                       \\
 --3 & 1/2 & $\Xi\Lambda$, $\Xi\Sigma$ & $\{27\}$, $\{8\}_s$            \\
     & 3/2 & $\Xi\Sigma$               & $\{27\}$                       \\
 --4 &   1 & $\Xi\Xi$                  & $\{27\}$
\end{tabular}
\label{tabirrep}
\end{table}

\begin{figure}
\caption{Effective ``potential'' $W$ in MeV for the $^1S_0$ partial
         wave in the $N\!N$, $\Sigma N$, $\Sigma\Sigma$, $\Xi\Sigma$,
         and $\Xi\Xi$ channels. The results are for models NSC97a
         and NSC97f; the other models show very similar behavior.}
\label{figW27}        %%% figure 1
\end{figure}

\begin{figure}
\caption{Prediction of the total cross section in mb: (a) for $nn$, $np$,
         and purely nuclear $pp$ scattering; and (b) for $pp$ scattering
         including the Coulomb-nuclear interference modification.
         The results are for NSC97a, but the other five NSC97 models
         give identical results.}
\label{figNNsgt}        %%% figure 2
\end{figure}

\begin{figure}
\caption{Prediction of the total cross section in mb for (a) $\Sigma^-n$,
         (b) $\Sigma^+p$ including the Coulomb-nuclear interference
         modification, (c) $\Lambda p$, and (d) $\Lambda n$ scattering.}
\label{figYNsgt}        %%% figure 3
\end{figure}

\begin{figure}
\caption{Prediction of the total cross section in mb for
         (a) $\Lambda\Lambda$, (b) $\Xi^-n$, (c) $\Xi^0p$, (d) purely
         nuclear $\Sigma^+\Sigma^+$, and (e) $\Sigma^+\Sigma^+$
         scattering including the Coulomb-nuclear interference
         modification. The nuclear and Coulomb-nuclear interference
         results for $\Sigma^-\Sigma^-$ scattering are practically
         identically to the corresponding $\Sigma^+\Sigma^+$ cases
         and, hence, are not shown explicitly.}
\label{figXNsgt}        %%% figure 4
\end{figure}

\begin{figure}
\caption{Prediction of the total cross section in mb for
         (a) $\Xi^0\Sigma^+$, (b) $\Xi^-\Sigma^-$ including the
         Coulomb-nuclear interference modification, (c) $\Xi^-\Lambda$,
         and (d) $\Xi^0\Lambda$ scattering.}
\label{figXYsgt}        %%% figure 5
\end{figure}

\begin{figure}
\caption{Prediction of the total cross section in mb for 
         (a) $\Xi^0\Xi^0$, (b) $\Xi^-\Xi^0$, (c) purely nuclear
         $\Xi^-\Xi^-$, and (d) $\Xi^-\Xi^-$ scattering including the
         Coulomb-nuclear interference modification.}
\label{figXXsgt}        %%% figure 6
\end{figure}

\end{document}